\documentclass[twocolumn]{aastex63}
\usepackage{enumitem}

\shorttitle{Decompression of neutron star tidal tails}
\shortauthors{C.J.~Nixon et.~al.}
\begin{document}
\title{Short gamma-ray bursts and the decompression of neutron star matter in tidal streams}
\author[0000-0002-2137-4146]{C.~J.~Nixon}
\affiliation{School of Physics and Astronomy, University of Leicester, Leicester, LE1 7RH, UK}
\author[0000-0003-3765-6401]{Eric~R.~Coughlin}
\affiliation{Department of Astrophysical Sciences, Princeton University, Princeton, NJ 08544, USA}
\affiliation{Department of Physics, Syracuse University, Syracuse, NY 13244, USA}
\author{J.~E.~Pringle}
\affiliation{School of Physics and Astronomy, University of Leicester, Leicester, LE1 7RH, UK}
\affiliation{Institute of Astronomy, Madingley Road, Cambridge, CB3 0HA, UK}

\email{cjn@leicester.ac.uk}

\begin{abstract}
Short gamma-ray bursts (sGRBs) are generally thought to result from the merger of two neutron stars or the merger of a neutron star with a black hole. It is becoming standard practise to model these mergers with hydrodynamical simulations that employ equations of state that are derived, for example, for determining the behaviour of matter in core-collapse supernovae (CCSNe), and which therefore make use of the assumption that the matter is hot and in nuclear statistical equilibrium (NSE). In this Letter we draw attention to the fact that the hydrodynamical timescale (roughly the gravitational timescale of the neutron star) may be several orders of magnitude shorter than the timescale on which such equilibrium can be re-established in the tidal debris ejected during a sGRB, and that on the initial decompression timescales the unshocked tidal ejecta may remain sufficiently cool that the employed equations of state are not appropriate for modelling the dynamics of this part of the flow. On timescales short compared with the timescale on which NSE can be (re)established, the equation of state can remain relatively stiff and thus the stream of tidal debris can remain narrow and vulnerable to gravitational instability, as has recently been suggested. These findings suggest that estimates of the type and abundances of heavy elements formed in short gamma-ray bursts need to be revisited. We suggest that the most direct method of testing the physical and dynamical properties of tidal ejecta in sGRBs will come from modelling of their light curves, which provides the cleanest source of information on the system dynamics.
\end{abstract}


\section{Introduction}
Short gamma-ray bursts (sGRBs) are thought to result from mergers of binary systems composed of two neutron stars or a neutron star and a black hole. The final stages of the merger consist of a plunge of a neutron star into close proximity of the other object, during which the neutron star is tidally disrupted. This process results in the production of one or more streams of tidal ejecta \citep[dynamical ejecta; see the review by][]{Shibata:2019aa}. Most, but not all, of this material falls back onto the remnant-disc system at subsequent times. This remnant-disc system can also give rise to the ejection of material. In this paper we concern ourselves with the tidally ejected material alone.

In a recent paper, \citet[][see also \citealt{Coughlin:2016ab,Coughlin:2016aa}]{Coughlin:2020ab} investigated the gravitational stability of such tidal streams, focussing on the case where a neutron star is disrupted by a black hole. They modelled the equation of state of the debris stream as being polytropic with $P = K \rho^\gamma$, where $P$ and $\rho$ are the pressure and density of the material comprising the stream, and $K$ and $\gamma$ are taken to be constant \citep[see also, e.g.,][]{Rasio:1992aa,Lee:1999ab,Lee:2000aa,Lee:2007aa,Xie:2020aa}. Using both analytical calculations (applicable to the general case) and supporting numerical simulations (in which they employed $\gamma=1.8-3$, concentrating on $\gamma=2$), \cite{Coughlin:2020ab} showed that the debris is unstable to the formation of knots along the stream provided $\gamma \ge 5/3$. Their simulations found that the debris stream comprised around 10 per cent of the original neutron star, the knots were separated by distances of order the width of the stream and had masses of order $0.01 M_\odot$, and the knots occupied both the bound and unbound portions of the stream. \cite{Coughlin:2020ab} noted that such behaviour can manifest itself as variability of the lightcurves of sGRBs \citep[see also][]{Colpi:1994aa}. Similar numerical results have also been reported by, for example, \cite{Rasio:1994aa,Lee:2000aa,Lee:2007aa}.

There are many different equations of state proposed for the matter that makes up static neutron stars, but while they differ in detail, they generally show the same trends \citep[see the reviews by][]{Lattimer:2001aa,Lattimer:2016aa,Ozel:2016aa}. Although the equation of state at neutron star densities ($\rho \gtrsim 10^{14}$\,g/cm$^3$) has an effective polytropic index of around $\gamma_{\rm eff} \approx 2.8 - 3.2$ \citep[see, e.g., Fig. 10 and Fig. 7 of][respectively]{Potekhin:2013aa,Ozel:2016aa}, at lower densities relevant to the stream of tidal debris the effective polytropic index drops rapidly. Indeed for the densities of $\rho \approx 10^{11-13}$\,g/cm$^3$ at which \cite{Coughlin:2020ab} find that gravitational instability has had time to set in, the effective polytropic index drops to $\gamma_{\rm eff} \lesssim 1.3$ \citep[see, for example, Fig.~10 in][]{Potekhin:2013aa}.\footnote{The stiffening of the equation of state at higher densities is attributed to a phase transition as the matter goes from an inhomogeneous mixture of nucleons, nuclear clusters, heavy nuclei and electrons, to a homogeneous ``bulk'' matter comprised mainly of neutrons \citep{Lattimer:1991aa,Lattimer:2016aa}. During core-collapse supernovae (CCSNe), for which the equations of state were designed, it seems reasonable to imagine that the high temperatures and vigorous gravitational collapse provide the necessary conditions for such a phase transition to occur rapidly. However, in the case of decompressing neutron star matter, it is clear that such a phase transition, operating in reverse, requires, for example, neutrons to decay into protons and electrons (the presence of, for example, the electrons is necessary to lower the effective adiabatic index, as their degeneracy provides additional pressure). The timescale for this to occur has not been fully addressed.} If this were the case, then as \cite{Coughlin:2020ab} show, gravitational instability of the stream would not occur.

Hydrodynamical modelling of neutron star mergers and the tidal debris resulting from such mergers has been performed with different equations of state. For example, some studies have employed a polytropic equation of state with $\gamma = 2$ \citep[e.g.\@][]{Rasio:1992aa,Lee:2007aa,Ruiz:2019aa,Coughlin:2020ab}. More sophisticated approaches use the equations of state for neutron star matter (e.g., the equations of state described in the reviews by \citealt{Lattimer:2001aa,Lattimer:2016aa,Ozel:2016aa}) to model both the neutron star structure and the tidal debris resulting from the merger \citep[e.g.\@][]{Rosswog:2002aa,Oechslin:2007aa,Kiuchi:2009aa,Rosswog:2013aa,Deaton:2013aa,Foucart:2014aa,Bernuzzi:2016aa,Radice:2016aa,Dexheimer:2019aa}. In these latter works, in contrast to (among others) the numerical simulations of \cite{Coughlin:2020ab} who assume $\gamma = 2$ throughout, the effective polytropic index $\gamma_{\rm eff}$ for the matter drops from around $\gamma_{\rm eff} \approx 3$ in the star to $\gamma_{\rm eff} \lesssim 1.3$ in the stream; and thus the pressure drop experienced by the debris is much smaller (for a given drop in density) and the streams are found to be much broader (see, for example, the ``enormous expansion'' shown in Figure 6 of \citealt{Rosswog:2002aa}). \cite{Lee:2007aa} comment that the fragmentation in tidal tails is not seen in merger simulations using ``realistic equations of state'', for which they refer to the calculations of \cite{Rosswog:2004aa} which employ the equilibrium equations of state derived by \cite{Shen:1998aa,Shen:1998ab}. Thus, unlike in the numerical simulations of \cite{Coughlin:2020ab}, which employed a polytropic equation of state with $\gamma =2$, the streams formed in simulations with equations of state that assume nuclear statistical equilibrium (NSE)\footnote{We note that a precise definition of ``NSE'' is not readily available. However, the conditions under which it is valid is discussed by \cite{Hix:1999aa}.} is readily established on the decompression timescale are found to be not susceptible to gravitational instability, as \cite{Coughlin:2020ab} predict for these equations of state (i.e., where $\gamma_{\rm eff} < 5/3$ for the debris).

In this paper, we draw attention to the fact that the typical timescale associated with the production of tidal debris from a merging neutron star is of order $\sim 10$ milliseconds. In contrast, the timescale on which nuclear equilibrium can be established in the dense neutron rich matter can be substantially longer \citep{Lattimer:1977aa,Meyer:1989aa,Colpi:1993aa}. It therefore follows that the use of the standard equations of state for static neutron star matter \citep[e.g.\@][]{Lattimer:1991aa,Lattimer:2001aa,Lattimer:2016aa,Ozel:2016aa}, which assume that NSE has already been established, is not appropriate for modelling the evolution of the tidal debris following the merger of two neutron stars or the merger of a neutron star with a black hole.\footnote{We note that these equations of state were principally derived for fast computation of the hydrodynamics in, for example, CCSNe. In this case, the temperatures are sufficiently high that NSE may be established (\citealt{Hix:1999aa} give $T > 3\times 10^9$\,K as a necessary, but not sufficient, condition for NSE). It seems plausible that the use of such equations of state for modelling CCSNe, or the merged remnant of a neutron star-neutron star merger (which is shocked to high temperatures), is appropriate and may provide an adequate description of the dynamics \citep[but see also][who comment ``adiabatic cooling on timescales of seconds can cause conditions to change more rapidly than NSE can follow'']{Hix:1999aa}. However, the conditions for NSE do not appear to be readily available in the tidal debris thrown off during the merger.} 

In Section~\ref{minmass} we draw attention to an earlier discussion of this problem where it was shown by \cite{Colpi:1993aa} that earlier claims---that neutron stars below the minimum stable mass might explode---may not be correct if allowance is made for the discrepancy between the dynamical timescale on which the matter decompresses and the $\beta$-decay timescales on which nuclear equilibrium can be (re)established. In Section~\ref{eosstream} we consider the equation of state for neutron star tidal debris. In Section~\ref{discussion} we summarise and discuss observational consequences.

\section{Decompression of neutron star matter}
\label{minmass}
A star with an equation of state that has a mean effective polytropic index $\gamma_{\rm eff}$ that satisfies $\gamma_{\rm eff} < 4/3$ is gravitationally unstable \citep[see, for example,][]{Cox:1980aa}. As we have noted above, for neutron star matter the effective polytropic index (of the equations of state that assume NSE) drops with density, and for low enough densities can fall below $\gamma_{\rm eff} = 4/3$. Therefore it follows that there is likely to be a minimum mass (i.e., a minimum density) for neutron stars. In this case, since one is looking for static, equilibrium configurations, it is appropriate to use a nuclear equilibrium equation of state. The minimum mass for a stable neutron star has been found to be $M_{\rm min} \approx 0.1 M_\odot$, with a corresponding radius $R_{\rm min} \approx 200$ km \citep[][see also \citealt{Shapiro:1983aa}]{Wang:1970aa,Cohen:1971aa,Baym:1971aa}. A simple way of understanding this phenomenon is to note that the minimum mass occurs when the energy available through $\beta$-decay ($\approx 0.78$\,MeV per neutron) of the available neutrons is comparable to the gravitational binding energy. For a mass of neutrons of $M_{\rm min} \approx 0.1 M_\odot$, the available energy is $\Delta E_\beta \approx 1.5 \times 10^{50}$\,erg, which is comparable to the gravitational binding energy $E_{\rm grav} \approx GM_{\rm min}^2/R_{\rm min}  \approx 1.3 \times 10^{50} (M_{\rm min}/0.1 M_\odot)^2(200\,{\rm km}/R_{\rm min})$\,erg.

It was originally suggested that neutron stars below the minimum mass would explode \citep{Page:1982aa}\footnote{We thank Martin Rees for drawing this to our attention.}. If so, this would be of relevance to the gravitational stability of tidal debris streams in sGRBs, as the instability discussed by \cite{Coughlin:2020ab} results in the debris stream fragmenting into knots that (for $\gamma=2$) were found to have masses of $\approx 0.01M_\odot$, i.e., below the minimum neutron star mass. A scenario in which a neutron star might be reduced to below the minimum mass is through mass transfer in a neutron star-black hole binary. Indeed, \cite{Blinnikov:1990aa} carried out numerical hydrodynamical simulations, using a nuclear equilibrium equation of state, of a low mass neutron star transferring mass to a black hole, and suggested that the resulting explosions might result in detectable bursts of high-energy radiation \citep[see also][]{Yudin:2020aa}.

However, a number of authors \citep{Lattimer:1977aa,Meyer:1989aa,Colpi:1989aa,Colpi:1993aa,Colpi:1994aa,Sumiyoshi:1998aa,Hix:1999aa} have pointed out that it is not appropriate to use an equation of state that assumes NSE in a physical situation where the hydrodynamical timescale (here the gravitational timescale of the neutron star) is several orders of magnitude shorter than the timescale for establishing NSE (roughly the timescale for $\beta$-decay). \cite{Colpi:1993aa} demonstrate that, starting with a neutron star in hydrostatic equilibrium below the minimum mass, nothing happens except a slow quasi-hydrostatic expansion (and evaporation) phase, lasting $\sim 10^3 - 10^5$ s, until a time at which significant energy release due to $\beta$-decay is able to occur (see also the calculations of the composition of decompressing neutron star matter provided by \citealt{Lattimer:1977aa,Meyer:1989aa}). What happens following the slow quasi-hydrostatic expansion must depend crucially on the timescale on which the energy generated by $\beta-$decay can be radiated away. \cite{Colpi:1993aa} conclude that at that point the star is dispersed to infinity in an explosive fashion.

Thus it has been well demonstrated that the equation of state of decompressing neutron star matter depends crucially on the rate at which that decompression occurs. We consider the relevance of this to the behaviour of neutron star tidal ejecta in Section~\ref{eosstream}.

\section{The equation of state of the tidal debris}
\label{eosstream}
It is generally found that the mass ejected in the tidal streams of neutron star-neutron star mergers is of the order of $0.1-1$ per cent of the mass of the neutron stars, while for neutron star-black hole mergers the mass ejected in the tails is around 10 per cent assuming that the black hole is of low enough mass to disrupt the neutron star and not simply swallow it whole \citep[e.g.][]{Shibata:2019aa}. From detailed models of the internal structure of neutron stars, the mass contained within the crust (at densities $\lesssim 2 \times 10^{14}$\,g/cm$^3$) is of the order of one per cent of the mass of the star \citep[see, for example,][]{Chamel:2008aa}. 

For neutron star-neutron star mergers it is possible that, in the cases where the lowest amounts of ejecta occur, the tidal ejecta are predominantly crust material, but in general there will be some (and perhaps a majority of) core material present. In contrast, for neutron star-black hole mergers, \cite{Coughlin:2020ab} find that the material in the tidal ejecta comes predominantly from the core of the neutron star rather than from the inner crust, and thus the debris is comprised of material with initial density higher than the nuclear saturation density ($\rho > \rho_{\rm sat} \approx 2.8 \times 10^{14}$\,g/cm$^3$; \citealt{Ozel:2016aa}) at which the matter may not be composed of nucleons alone but may contain a rich variety of hadronic degrees of freedom \citep{Ostgaard:2001aa,Chamel:2008aa,Ozel:2016aa}. It is at densities of $\rho \approx \rho_{\rm sat}$ that neutron star matter undergoes a phase transition from consisting of ``inner crust'' material (predominantly a neutron gas with some nuclei that consist of proton clusters with small neutron fraction) to ``core'' material (a much less compressible neutron fluid, with nuclear forces dominated by the repulsive core of the nuclear potential).

In most of the recent literature it is generally assumed that the material ejected in the merger is initially in, and subsequently remains close to, NSE. For example, \cite{Shibata:2005aa,Roberts:2011aa,Foucart:2014aa,Shibagaki:2016aa,Brege:2018aa,Radice:2018aa} all use equations of state based on the work of \cite{Lattimer:1991aa}, who assume equilibrium with respect to strong and electromagnetic interactions, but explicitly do not assume $\beta$-equilibrium on the grounds that equilibrium with respect to weak interactions is often not achieved within the timescales of many astrophysical phenomena\footnote{However, as noted above, it is implicit in the \cite{Lattimer:1991aa} equation of state that the matter has time to undergo a phase transition from inhomogeneous to homogeneous matter, and this process requires time to allow neutrons to decay into protons and electrons.}. For NSE to be a valid assumption, \cite{Hix:1999aa} note that this requires endoergic reactions of each reaction pair to occur, and that a necessary, but not sufficient, condition for this to occur is for the temperature to be $T > 3 \times 10^9$ K. \cite{Thielemann:2017aa} comment that at temperatures $T < 3 \times 10^9$ K all nuclear reactions have to be followed in detail. \cite{Hix:1999aa} also note that in explosive Si burning, adiabatic changes occur on a timescale of seconds and that this causes changes in conditions to occur faster than NSE can follow. They emphasise further that  in the ``face of sufficiently rapid thermodynamic variation, NSE provides a problematic estimate of abundances'', and that, in addition, there are a number of astrophysically important situations where NSE is not globally valid including the decompression of neutron star matter.

Computations of the make-up of the decompressing tidal debris have been carried out by \citet[][see also \citealt{Lattimer:1977aa}]{Meyer:1989aa}. The computations presented by \cite{Meyer:1989aa} start with crust material in nuclear equilibrium at densities around $10^{12} - 10^{13}$\,g/cm$^3$, which are above the neutron drip density of $\sim 4 \times 10^{11}$ g/cm$^3$. The assumption is that the original pre-interaction neutron star material is cold \citep[][quoting \citealt{Bahcall:1965aa,Bahcall:1965ab}, notes that a lone neutron star cools to essentially zero temperature within a million years]{Meyer:1989aa} and that during the merger the material that is expelled in the tidal arms has undergone adiabatic expansion but has not undergone a shock. In their NS-BH merger computations, \cite{Coughlin:2020aa} find that this is true for the material that comprises the tidal streams.~\footnote{The same is found by \cite{Foucart:2014aa} in their simulations of black hole-neutron star mergers, and they note that the tidal tails consist of cool unbound neutron rich material. In this context, ``cool'' refers to temperatures of around $T \sim 4 - 10 \times 10^9$ K. How this tidally ejected material acquires such high temperatures is not considered, but it seems likely that it is due to a combination of the assumed equation of state and numerical effects.} \cite{Meyer:1989aa} notes that from the work of \cite{Lattimer:1977aa} it is to be expected that such initially neutron-rich material starts by forming neutron-rich nuclei, and that these emit neutrons as the matter expands and becomes less dense. Once the expansion rate has fallen enough, and the neutron density dropped sufficiently, $\beta$-decays can occur and heat the matter, eventually, to r-process temperatures.

\cite{Meyer:1989aa} assumes that decompression occurs with an expansion timescale $\tau_s$, defined as $\alpha$ times the local dynamical timescale; thus  
\begin{equation}
  \label{defalpha}
  \rho/|{\dot \rho}| = \tau_s = 446 \alpha \rho^{-1/2}\,{\rm s}\, , 
\end{equation}
where $\rho$ is the mass density in cgs units.
\citet[][see Figs 3--10]{Meyer:1989aa} finds, in line with expectations, that the expanding material becomes heated by $\beta-$decays only when the decay timescale becomes shorter than the expansion timescale. For $\alpha = 10$ (relevant to the simulations in \citealt{Coughlin:2020ab}, see below) he finds that by the time the density has decreased to $\approx 3\times10^{11}$\,g/cm$^3$ the temperature has not yet reached 0.08 MeV. We note that these temperatures are still insufficient for the validity of the assumption of NSE, which requires at least $\sim 0.25$ MeV \citep{Hix:1999aa}.

More recent calculations along the lines of those by \cite{Meyer:1989aa} are presented by \citet[][see \citealt{Arnould:2007aa} for discussion]{Goriely:2005aa}. They present the evolution of matter starting at $10^{14}$\,g/cm$^3$ and expanding on a timescale of $6.5$\,ms. The matter is initially assumed to be at a temperature of $T = 10^8$\,K, and is found to remain at this temperature, with no evolution of the nuclei, until a time of around $80$\,ms -- by this time \cite{Coughlin:2020aa} find that the stream is already fragmented into knots.

In contrast to these works, more recent computations by, for example, \cite{Lippuner:2015aa}, aimed at predicting the nuclear abundances in stripped material, typically start with material in NSE at lower densities ($\rho \approx 10^6 - 10^{12}$\,g/cm$^3$) and already higher temperatures ($T_0 = 6 \times 10^9\,{\rm K} \approx 0.52$\,MeV). Similarly \cite{Roberts:2017aa} consider the late evolution of the tidal tails found by \cite{Foucart:2014aa}. The tail material has already decompressed considerably ($\rho < 10^{12}$ g/cm$^3$) but, for reasons that are unclear, is already hot enough ($T > 10^{10}$ K) for NSE equations of state to be applicable.

Finally, we note that most of the material that ends up in the tidal streams in the NS-BH merger simulations of \cite{Coughlin:2020ab} is initially core material. The equation of state, and even the composition of the matter, at these high densities is still a matter of debate \citep{Lattimer:1991aa,Ostgaard:2001aa,Lattimer:2001aa,Ozel:2016aa,Greif:2020aa} with uncertainties in the pressure at a given density being around a factor of five (e.g., the left panel of Fig.~7 of \citealt{Ozel:2016aa}). However, there is general agreement that the equation of state at these densities is stiff, with the local polytropic index being $\approx 3$ \citep[e.g., Fig. 10 of][]{Potekhin:2013aa}. Thus, {\em a fortiori}, it follows that the equation of state of such material that is decompressed to form tidal streams on timescales of order milliseconds is essentially unknown. In \cite{Coughlin:2020ab} it is found that the material that forms the self-gravitating knots starts at densities of around $5 \times 10^{14}$\,g/cm$^3$ and expands rapidly. The knots then form in the stream by the time the density has reduced to around $\rho \approx 3 \times 10^{11}$\,g/cm$^3$ for $\gamma = 2$ and to around $\rho \approx 2\times 10^{13}$\,g/cm$^3$ for $\gamma = 3$. The timescale on which this decompression occurs corresponds approximately to Eq.~\ref{defalpha} with $\alpha \approx 10-20$ for $\gamma=2$ and $\alpha \approx 30-60$ for $\gamma=3$. The time after which the debris stream has formed substantial self-gravitating knots is $\approx 50$\,ms for $\gamma=2$ and $\approx 10$\,ms for $\gamma=3$. We note that the densities at which the knots have already formed in the stream, particularly for $\gamma=3$, are of the order of, or higher than, the densities at which e.g. \cite{Lattimer:1977aa} and \cite{Meyer:1989aa} begin their calculations for the evolution of the decompressing matter.

\section{Discussion}
\label{discussion}
The debris ejected as tidal streams from neutron star-neutron star and neutron star-black hole mergers is initially unshocked and therefore cool, and typically decompresses on a timescale much shorter than the timescale on which it can achieve NSE. We reiterate that the decompression of neutron star matter in tidal streams is not equivalent to the compression of matter in CCSNe; the temperature of the matter is very different and no longer justifies the assumption of NSE and the matter is not subject to strong gravitational contraction that may provide the necessary conditions for a rapid phase transition in the matter between inhomogeneous and homogeneous. Thus the use of the standard neutron star equations of state for the tidal debris, which are based on the assumption of NSE, is not appropriate. In particular such an assumption leads to streams that are much broader and of much lower density than should be the case. This has a number of implications:

\begin{enumerate}
\item Narrow tidal streams are more susceptible to self-gravity, and so to the formation of knots which can lead to variability and potentially flaring events in the light-curves of sGRBs \citep{Colpi:1994aa,Coughlin:2020ab}.

\item Any such knots that escape the system are likely to evolve, mainly through $\beta$-decay. If the energy released from $\beta$-decay is retained, then these objects may explode \citep{Colpi:1993aa}, which may provide an alternative source of variability in the light-curves of sGRBs \citep{Colpi:1994aa}. If, however, a significant fraction of the energy released can be efficiently radiated away, then the knots may evolve into high-velocity planetary mass ($\sim 0.01 M_\odot$) objects of peculiar composition.

\item Estimates of the type and abundances of high-Z nuclei, for example gold \citep{Rosswog:2002aa} and lanthanides \citep{Metzger:2010aa,Metzger:2015aa,Lippuner:2015aa} from the tidal ejecta may need to be revisited, with more realistic initial conditions \citep[cf.\@ the calculations of][]{Lattimer:1977aa,Meyer:1989aa}. These calculations often begin with matter that is hot and has densities that are appropriate to neutron star crust material. Such densities may represent the primary ejecta in neutron star-neutron star mergers, but these initial densities are too low compared with the initial densities of the tidal ejecta expected from neutron star-black hole mergers.

\item As the amounts, and the initial densities, of matter ejected in tidal tails from neutron star-neutron star mergers are generally found to be smaller than that found in neutron star-black hole mergers, it is possible that the gravitational stability properties and the nuclear abundances in these events are sufficiently different that we might expect different observable properties in these two populations of sGRB progenitors.
\end{enumerate}

\subsection{Where next?}
The problems we have drawn attention to in this Letter have at their heart the physical characteristics and properties of the unbound portions of the tidal streams that result from the merger of two compact objects (neutron star-neutron star or neutron star-black hole). Much of the interest in this area comes from predicting the nuclear composition of this material. For example, \cite{Radice:2018aa} note that ``most calculations of nucleosynthesis~\ldots~involve taking density histories, $\rho(t)$, of Lagrangian tracers and evolving the composition and entropy of the material in time starting [with initial] entropy and electron fraction extracted from the simulation output''.

Early work in this area \citep[for example][]{Rosswog:1999aa,Freiburghaus:1999aa} used Lagrangian hydrodynamics (SPH) and Newtonian gravity together with an equation of state for hot and dense nuclear matter, based on \cite{Lattimer:1991aa}. As mentioned above, because such equations of state become highly compressible once the density drops below around $\rho \approx 10^{14}$ g/cm$^3$ \citep[e.g.\@][]{Potekhin:2013aa}, the tidal arms produced in the simulations tend to be broad and low density. In the most recent simulations, attention has switched to obtaining gravitational wave-train predictions for such mergers. Thus, for example, \cite{Foucart:2014aa} use pseudo-spectral methods to compute the evolution of the gravitational metric hybridised with a finite difference code to compute the hydrodynamics \citep[for details see, for example,][]{Haas:2016aa}. They use the \cite{Lattimer:1991aa} equation of state for nuclear material. However, in \cite{Foucart:2014aa}, the few per cent of the total fluid that ends up in the unbound portion of the tidal tails is not well resolved. Given that the kinetic velocities of the ejecta is acknowledged to be resolution dependent, it seems likely that estimates of the entropy (which imply $T \sim 4 - 8 \times 10^9$\,K) and density ($\rho \sim 3 \times 10^{11}$\,g/cm$^3$) of the ejected material \citep[][Figure 4]{Foucart:2014aa} are unreliable. The properties of the small fraction of the initial stellar material that ends up as unbound tidal ejecta is then used as the starting point for nucleosynthesis computations \citep[e.g.\@][]{Roberts:2017aa}.

In contrast to the above, \cite{Lattimer:1977aa} and \cite{Meyer:1989aa} argue that a realistic expectation for the computation of nucleosynthesis in decompressing neutron star matter would be to assume that initially $T \approx 0$\,K. This would be appropriate for an old (age $> 10^6$ yr) neutron star which has had time to cool \citep{Bahcall:1965aa,Bahcall:1965ab}, and for ejecta that are subject simply to tidal forces and not shocks. Such material is cold, dense and subject to self-gravity \citep{Colpi:1994aa,Coughlin:2020ab}.

It is evident that starting nucleosynthesis calculations from such different initial conditions is likely to lead to quite different conclusions. Thus the most fruitful initial step towards resolving this tension is to find some way of computing in a credible manner the initial properties of the tidal ejecta.

We finish by noting that information about the system dynamics is most directly obtained from the time-dependence of the light curves. Much of the recent work in this area has been aimed at determining the abundances of heavy elements formed in such mergers, and as such has been less focussed on understanding the {\em dynamics} of the tidal tails. As we have seen, the time-dependence of the properties of the nuclear matter (often treated simply as an equation of state) has clear consequences for the physical properties and dynamical behaviour of the tidal streams. We argue that the difference between the tidal streams being narrow, high density and clumped and being broad, low density and smooth is likely to show itself more clearly in the time-dependence of the rate of energy release from the fallback material. This is discussed by \cite{Colpi:1994aa,Coughlin:2020ab}. Thus we suggest that identifying models that are capable of explaining the light curve properties, and in particular the detailed variability properties, is perhaps the most convincing way of making progress in understanding the properties of the nuclear physics at these densities and temperatures.

\acknowledgments
We thank Martin Rees for stimulating our interest in the details of the nuclear equation of state. CJN is supported by the Science and Technology Facilities Council (grant number ST/M005917/1), and funding from the European Union’s Horizon 2020 research and innovation program under the Marie Sk\l{}odowska-Curie grant agreement No 823823 (Dustbusters RISE project). ERC acknowledges support from NASA through the Hubble Fellowship, grant No. HST-HF2-51433.001-A awarded by the Space Telescope Science Institute, which is operated by the Association of Universities for Research in Astronomy, Incorporated, under NASA contract NAS5-265555.

\bibliographystyle{aasjournal}
\bibliography{nixon}

\end{document}